\documentclass[aps,amsmath,amssymb,reprint,longbibliography]{revtex4-1}

\usepackage{graphicx}
\usepackage{amsmath,amssymb}
\usepackage{textcomp}
\usepackage{hyperref}
\usepackage{xcolor}
\usepackage{bm}

\newcommand{\vns}{\ensuremath{v_\mathrm{ns}}}

\newcommand{\bv}[1]{\ensuremath{\mathbf{#1}}}

\def\3He{$^3$He}
\def\4He{$^4$He}

\begin{document}

\title{Temporal decay of vortex line density in rotating thermal counterflow of He~II }

\author{F. Novotný}
\author{M. Talíř}
\author{E. Varga}\email{emil.varga@matfyz.cuni.cz}
\author{L. Skrbek}\email{ladislav.skrbek@matfyz.cuni.cz}
\affiliation{Faculty of Mathematics and Physics, Charles University, Ke Karlovu 3, Prague, 121 16, Czech Republic}

\date{\today}

\begin{abstract}
 Horizontally ($\mathbf{\Omega} \perp \mathbf{v}_{\rm{ns}}$) and axially ($\mathbf{\Omega} \parallel \mathbf{v}_{\rm{ns}}$) rotating counterflow of superfluid $^4$He (He~II) generated thermally in a square channel is studied using the second sound attenuation technique, detecting statistically steady state and temporal decay of the density of quantized vortex lines $L(t,\Omega)$. The array of rectilinear quantized vortices created by rotation at angular velocity $\Omega$ strongly affects the transient regimes of quantum turbulence characterized by counterflow velocity $\mathbf{v}_{\rm{ns}}$, differently in both geometries. Two effects are observed, acting against each other and affecting the late temporal decay $L(t,\Omega)$. The first is gradual decrease of the decay exponent $\mu$ of the power law $L(t,\Omega) \propto t^{-\mu}$, associated with the fact that under rotation thermal counterflow acquires two-dimensional features, clearly observed and recently reported by us (Phys. Fluids \textbf{36}, 105121 (2024)) in the $\mathbf{\Omega} \parallel \mathbf{v}_{\rm{ns}}$ geometry. It exists in the $\mathbf{\Omega} \perp \mathbf{v}_{\rm{ns}}$ geometry as well, however, it is screened here by the influence of the effective Ekman layer built within the effective Ekman time 
 of order seconds. 
For faster rotation rates $L(t,\Omega)$ gradually ceases to display a clear power law. Instead, rounded and ever steeper decays occur, gradually shifted toward shorter and shorter times, significantly shortening the time range for a possible self-similar decay of vortex line density. This effect is not observed in $\mathbf{\Omega} \parallel \mathbf{v}_{\rm{ns}}$ geometry, as here the much longer effective Ekman time 
 of order minutes cannot affect the observed $L(t,\Omega)$ decay appreciably.

\end{abstract}

\maketitle
\section{Introduction}

Since the pioneering experiments and theoretical considerations of Vinen \cite{Vinen1957a,Vinen1957b,Vinen1957,Vinen1958}, thermal counterflow of superfluid $^4$He (He~II) 
represents the most studied type of quantum turbulence \cite{QTbook} in He~II. It can be easily set by applying heat flux $\dot q$ to a heater at a closed end of a channel, which is carried away in a convective manner by the normal fluid. By conservation of mass, a superfluid current arises in the opposite direction, and counterflow velocity is established: $\vns = \dot q/ \rho_\mathrm{s} s T$, where $T$ is the temperature and $s$ denotes the specific entropy. For sufficiently low $\vns$, the flow of the viscous normal fluid is laminar, and there are no quantized vortices (except for the remnant ones \cite{Awschalom1984}) in the potential flow of the superfluid component. Upon increasing $\dot q$, on exceeding the small critical counterflow velocity $\vns^\mathrm{cr}$, thermal counterflow becomes turbulent -- a tangle of quantized vortices is generated by extrinsic nucleation and reconnections. Both the superfluid component (as a tangle of quantized vortices) and the viscous normal fluid can become turbulent \cite{tough_rev,Marakov2014,Gao2017}, each having its own velocity field, which are coupled by the mutual friction force. One speaks about the unique, double turbulence regime. Despite the long history of its investigation, some of rich physical properties of thermal counterflow, such as scenarios of generation \cite{Skrbek2024PNAS} and temporal decay \cite{Gao2016} have been discovered only recently. 

He~II is a quantum liquid with extraordinary properties that cannot by described classically, by Navier-Stokes equation. Above $\approx 1$~K and at low Mach numbers \cite{Cetin2025}, its flow can be viewed in the frame of the phenomenological two-fluid model as a superposition of viscous normal flow (with temperature-dependent density $\rho_{\rm n}$ and velocity $\bv v_{\rm n}$) and inviscid superflow  of the superfluid component ($\rho_{\rm s}$, $\bv v_{\rm s}$). It can be described with a macroscopic wave function $\Psi$ with a macroscopically coherent phase $\Phi$ that severely restricts its velocity field to $\bv v_{\rm s}= \hbar/m_4\nabla \Phi$. The superflow is thus potential, however, vorticity is possible in the form of quantized vortices, line-like topological defects with diameter $\xi \approx 0.15$~nm around which the flow circulation is restricted to a single quantum of circulation $\kappa\approx 9.98\times 10^{-8}$ m$^2$/s \cite{Tilley_book,QTbook}. In the steady-state under rotation, the superfluid sample of He~II becomes threaded by a hexagonal lattice of rectilinear singly quantized vortex lines \cite{Osborne1950}, and on length scales larger than the mean distance between quantized vortices, He~II mimics classical solid body rotation. The vortex line density (total length of vortices per unit volume or, equivalently, the number of rectilinear vortices per unit area) obeys the Feynman rule
\begin{equation}
    L =\frac{2\Omega}{\kappa},
    \label{eq:Feynman}
\end{equation}
where $\Omega$ is the angular velocity of rotation.

Rotation significantly modifies classical as well as quantum turbulent flows. Rotating classical turbulent flows, rather than being driven primarily by the interaction with the boundaries, are strongly affected by the Coriolis body force $\bv F_c = -2\rho\boldsymbol{\Omega}\times \bv u$, where $\boldsymbol{\Omega}$ is the angular velocity, $\bv u$ denotes the velocity field and $\rho$ is the density. In rapidly rotating flows, characterized by low Rossby number $\mathrm{Ro} = u/l\Omega$ (with $u$, $l$ being characteristic velocity and length scale, respectively), the competition between excitation of inertial waves \cite{Bewley2007} 
and the two-dimensionalization of the flow due to the Taylor-Proudman theorem \cite{Davidson2015} leads to a phase transition-like appearance of the inverse cascade \cite{Kan2020,Kan2022a} which joins a growing body of intensely-studied phase-transition-like phenomena in fluid turbulence \cite{Cortet2010,Hof2023,Novotny2025}. The inverse cascade can lead to the spontaneous formation of large-scale structures \cite{Boffetta2012} or an intermediate regime with an energy flux loop \cite{ClarkDiLeoni2020} with the inverse cascade arrested at an intermediate scale. 

Rotating vortical quantum flows display additional interesting phenomena. One example is the rotating bucket of He~II, where the hexagonal array of vortex lines can become unstable in counterflow oriented along the axis of rotation \cite{Cheng1973,Swanson1983,Peretti2023} -- the so-called Donnelly-Glaberson instability \cite{Glaberson1974,Ostermeier1975} -- which leads to the excitation of helical Kelvin waves propagating along the axis of rotation \cite{Glaberson1974}, with the dispersion relation 
\begin{equation}
    \label{eq:dispersion}
    \omega = 2\Omega + \beta k^2\,,
\end{equation}
where $\beta = \kappa/4\pi \ln\left(b/\xi\right)$ and the wave-vector $\bv k$ is assumed to be oriented parallel to the axis of rotation; $b$ is the typical distance between quantized vortices. We note that a similar instability is believed to be at least partly responsible for the transition to turbulence also in non-rotating thermal counterflow through the so-called vortex mill mechanism \cite{Schwarz1990}. Experiments of Swanson \textit{et al.} \cite{Swanson1983} in rotating counterflow at small velocities and vortex filament numerical simulations of Tsubota \textit{et al.} \cite{Tsubota2003,Tsubota2004} showed the suppression of the critical velocity for the turbulent growth of vorticity to very small values for rapidly rotating turbulence and possibly a second critical velocity connected to the development of disordered tangle. The excitation of vortex waves was directly visualized by Peretti \textit{et al.} \cite{Peretti2023} who, however, forced the waves not by Donnelly-Glaberson instability but rather by oscillatory thermal counterflow. 

The vortex line density in rotating counterflow, $L(v_{\rm{ns}}, \Omega)$, is generated upon exceeding the critical velocity $v_{\rm{ns}}^{\rm{cr}}(\Omega)$ and can be determined from the frequency sweeps of the second sound resonance, as shown in Fig.~2 of our previous work \cite{Dwivedi}, where we experimentally studied rotating thermal counterflow in the parallel geometry of the rotation axis with the direction of thermal counterflow, ($\mathbf{\Omega} \parallel \mathbf{v}_{\rm{ns}}$). Our experimental data \cite{Dwivedi}  confirm strong suppression of the critical velocity $v_{\rm{ns}}^{\rm{cr}}(\Omega)$ for the turbulent growth of vorticity to very small values for rapidly rotating turbulence. 
Additionally, we extended the results of Swanson \textit{et al.} \cite{Swanson1983} to the dynamics of turbulence growth and decay. By performing complementary numerical simulations, we related the initial growth of vortex line density, $L(t,\Omega)$, to Kelvin wave growth on individual quantized vortices in the vortex lattice and showed that rotating quantum turbulence decays quasiclassically at long times consistent with rotating classical turbulence in elongated domains \cite{Morize2006,Kan2020,QTbook}. 

In this work, we extend our investigations to the perpendicular geometry: the horizontal counterflow is rotated about the vertical axis ($\mathbf{\Omega} \perp \mathbf{v}_{\rm{ns}}$). Following this introduction, we describe our experimental setup 
and present the obtained experimental results. We then discuss them in comparison with the results obtained in parallel geometry \cite{Dwivedi}, 
before summarizing our findings and drawing conclusions.

\section{Experimental setup}
\label{sec:setup}

\begin{figure}[h]
    \centering
    \includegraphics[width=.99\linewidth]{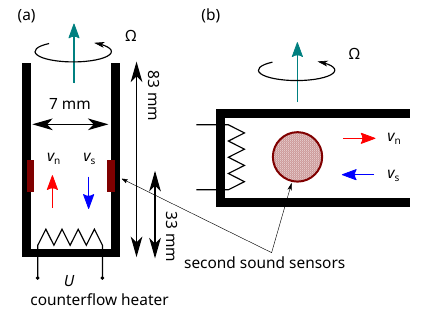}
    \caption{Schematic diagram of the $7\times 7$~mm$^2$ square cross-section counterflow channel, as used in experiments on rotating channel counterflow in a) parallel ($\mathbf{\Omega} \parallel \mathbf{v}_{\rm{ns}}$) and b) perpendicular ($\mathbf{\Omega} \perp \mathbf{v}_{\rm{ns}}$) geometries. The resistive heater placed in its dead end generates the counterflow of normal and superfluid components. Vortex line density is measured using second sound attenuation. }
    \label{fig:setup}
\end{figure}

The geometry of the counterflow channel used in our experiments is illustrated in Fig.~\ref{fig:setup}. In the previous experiment \cite{Dwivedi}, in $\mathbf{\Omega} \parallel \mathbf{v}_{\rm{ns}}$ geometry, it was nearly identical to the past experiments of Swanson \emph{et al.} \cite{Swanson1983}. In both  $\mathbf{\Omega} \parallel \mathbf{v}_{\rm{ns}}$ and  $\mathbf{\Omega} \perp \mathbf{v}_{\rm{ns}}$ geometries, the same brass flow channel with a square cross-section $A = 7 \times 7$~mm$^{2}$ and length $H = 83$~mm has been used, placed vertically and horizontally with respect to the axis of rotation of the cryostat, which is placed on a custom-made rotating platform that also supports all necessary electronic equipment (for details and a photograph, see Ref. \cite{Novotny2024}). The platform ensures steady rotation as well as  acceleration/deceleration between rest and the maximum velocity of 60 RPM in less than a second. A resistive wire heater ($R \approx 16$~\textohm) glued to a flat surface is inserted inside the channel from the bottom to provide the counterflow heat flux, keeping the other end to the liquid helium bath open.

The vortex line density in the channel is detected using second sound attenuation \cite{Varga2019}. As in \cite{Dwivedi}, the channel acts as a semi-open acoustic resonator for the second sound, which is excited and detected by a pair of oscillating membrane ($<1$~\textmu m pore size) second sound transducers \cite{Zimmermann1986}. The second sound sensors are placed symmetrically around the axis of rotation such that the region where vortex line density is detected lies on the axis of rotation. The change in the attenuation of the resonance is used to calculate the vortex line density $L$. The attenuation due to quantized vortices depends on the angle $\theta$ between the vortex and direction of sound propagation as $\sin^2\theta$, i.e., the second sound is not dissipated when propagating parallel to a straight vortex \cite{Babuin2012}.
For a disordered, turbulent tangle, one can assume a random orientation of quantized vortices uniformly distributed in all directions, which results \cite{Babuin2012} in

\begin{subequations}
    \label{eq:L-both}
    \begin{equation}
        \label{eq:L-random}
        L =\frac{6\pi\Delta_0}{B\kappa}\left(\frac{a_0}{a}-1\right)\,;
    \end{equation}
where $B$ is a temperature-dependent mutual friction parameter tabulated in Ref. \cite{Donnelly1998}, $\Delta_0$ is the width of the unattenuated second sound resonance, and $a_0$, $a$ are the amplitudes on resonance of unattenuated and attenuated peaks, respectively.

If all vortices are oriented perpendicular to the direction of sound propagation  then $\sin^2\theta = 1$ and the expression for vortex line density is instead
   
    \begin{equation}
        \label{eq:L-polarized}
        L =\frac{4\pi\Delta_0}{B\kappa}\left(\frac{a_0}{a}-1\right) \,,
    \end{equation}
\end{subequations}

This expression is appropriate for the case of steady rotation without counterflow, however, we have not used it in the present experiment, which is focused on the temporal decay of high values of vortex line density. Generally, since the polarization of the tangle cannot be determined, one should take into account both the statistical uncertainties and an additional systematic uncertainty up to 33\%. This complicates quantitative analysis of competition between rotation and counterflow in terms of $L(\mathbf{v}_{\rm{ns}}, \Omega)$.  

As argued in Ref.~\cite{Dwivedi}, the typical kinetic and thermal response time \cite{Varga2018} of the present experiment are of the order of 10~ms, and second sound response time (inverse linewidth of the resonance) is, at most, approximately 40~ms. This ensures that the transient effects studied, which occur on the timescale of seconds, are significantly affected neither by  relaxation times of the non-turbulent system nor by  relaxation times of the measuring electronics.

\begin{figure}[t]
    \includegraphics{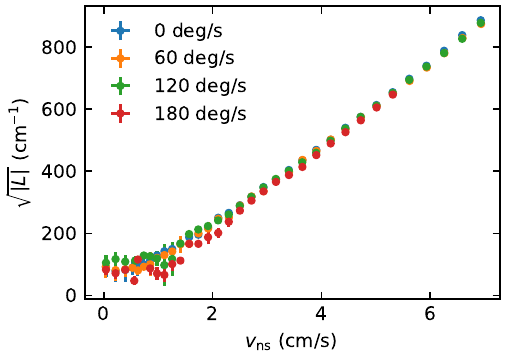}
    \caption{Square root of vortex line density in steady state counterflow turbulence, $\sqrt{L(v_{\rm{ns}},\Omega)}$, rotating in the $\mathbf{\Omega} \perp \mathbf{v}_{\rm{ns}}$ geometry at temperature 1.65 K, plotted as a function of counterflow velocity for several rotation speeds of the cryostat as indicated. 
    }
    \label{fig:steady-state}
\end{figure}

\section{Experimental results}

In view that the two experiments are performed in the same counterflow channel and differ only in the direction of rotation axis with respect to the direction of thermal counterflow, for clarity we describe the results in both geometries and directly compare them.

\subsection{Steadily rotating counterflow}

In both geometries we verified the performance of our experimental setup.  For $\mathbf{\Omega} \parallel \mathbf{v}_{ns}$ case we first measured the vortex line density in steady rotation and confirmed that $L$ obeys Feynman's rule, Eq.~\eqref{eq:Feynman}, see Fig.~1 of Ref.~\cite{Dwivedi}. The vortex line density in rotating counterflow, $L(v_{ns}, \Omega)$, determined from the frequency sweeps of the second sound resonance has been shown in Fig.~2 of Ref.~\cite{Dwivedi}; in the $\mathbf{\Omega} \perp \mathbf{v}_{\rm{ns}}$ case the result is similar; as shown in  Fig.~\ref{fig:steady-state}. 

Without counterflow, the situation in both geometries is essentially the same in that all quantized vortices are expected to be aligned parallel with the axis of rotation and perpendicular to the direction of propagation of the second sound, therefore vortex line density $L$ calculated using \eqref{eq:L-polarized} has to be the same.

Under rotation, in agreement with Swanson \textit{et al.}~\cite{Swanson1983}, the vortex-free state is suppressed as the rotation rate increases, which we now confirm for both $\mathbf{\Omega} \parallel \mathbf{v}_{\rm{ns}}$ and $\mathbf{\Omega} \perp \mathbf{v}_{\rm{ns}}$ geometries. For high enough counterflow velocities, in both $\mathbf{\Omega} \parallel \mathbf{v}_{\rm{ns}}$ \cite{Dwivedi} and $\mathbf{\Omega} \perp \mathbf{v}_{\rm{ns}}$ (see Fig.~\ref{fig:steady-state}) the expected $\sqrt{L(v_{\rm{ns}},\Omega)} \propto v_{\rm{ns}}$ scaling is reproduced, which occurs in flows with forced nonzero $v_{\rm{ns}}$ regardless of the angular velocity of the system.

\subsection{Temporal decay of rotating counterflow in $\mathbf{\Omega} \perp \mathbf{v}_{ns}$ and $\mathbf{\Omega} \parallel \mathbf{v}_{ns}$ geometries}

\begin{figure}
    \centering
    \includegraphics{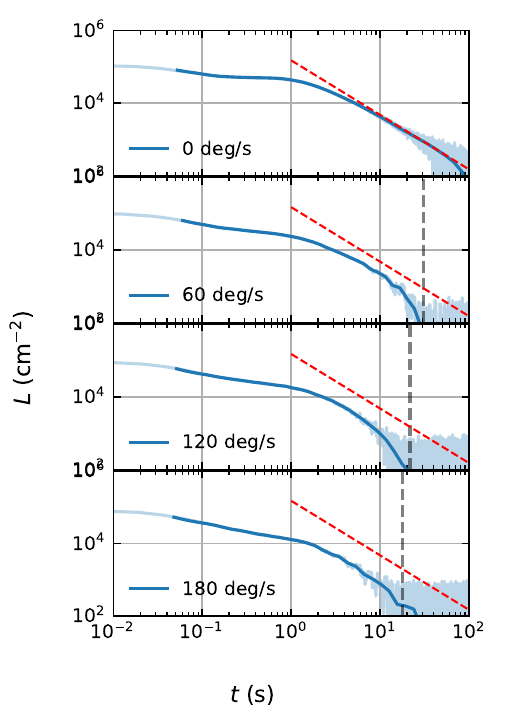}
    \caption{Temporal decay of the vortex line density $L(t, \Omega)$ for     
    rotating thermal counterflow in $\mathbf{\Omega} \perp \mathbf{v}_{\rm{ns}}$ geometry. The red dashed lines represent identical power law decays with the classical decay exponent $\mu=1.5$ \cite{Stalp1999,Morize2006}. The vertical dashed lines indicate the effective Ekman times  $T_{\rm{Ek}}^{\rm{eff}}=H (\nu_{{\rm{eff}}} \Omega)^{-1/2}$, where $\nu_{\rm{eff}} =\kappa/2$ is used as the value for the effective kinematic viscosity of turbulent He~II. This gives effective Ekman times 30.6 s, 21.6 s and 17.6 s for rotation rates 60, 120 and 180 degrees per second. }
    \label{fig:decayCompar}
\end{figure}

Fig.~\ref{fig:decayCompar} shows several examples of decaying vortex line density $L(t, \Omega)$ generated thermally under steady horizontal rotation of the channel ($\mathbf{\Omega} \perp \mathbf{v}_{\rm{ns}}$) when, at the time instant $t=0$, the heater is switched off. The decay thus does not proceed to zero (remnant) vortex line density, but to the equilibrium value given by the Feynman rule, Eq.~(\ref{eq:Feynman}), for the particular rotation rate. However, since the second sound attenuation technique measures the excess of vortex line density with respect to a reference state (in this case the steady rotation), the \emph{measured} $L(t)$ relaxes toward 0. The decay curves, masured here at $T=1.35$~K under steady rotation rates as indicated are averaged over multiple realizations (typically 50) and filtered (Savitzky–Golay filtering, window size 21, and polynomial order 3). The filtered data are indicated with the darker colors and the light-colored curves in the background show unfiltered data. 

The typical feature of all decay curves is the appearance of a bump at times past about 1 s of the decay, occurring in temporal decays $L(t,\Omega)$ both without and under rotation, for both $\mathbf{\Omega} \perp \mathbf{v}_{\rm{ns}}$ and $\mathbf{\Omega} \parallel \mathbf{v}_{\rm{ns}}$ (see Fig. 8 of Ref.~\cite{Dwivedi}) geometries.
The form of the decay curve at rest is typical as observed in previous experiments and discussed in Refs.~\cite{Babuin2016,Gao2016}. Over a wide range of temperatures the part of the decay that follows the bump takes the form of a power law with the decay exponent $\mu=3/2$ at late times; the decays differ only in the prefactor which serves to evaluate the effective kinematic viscosity of turbulent He~II \cite{Stalp1999}. In our previous study we have shown that steady rotation in parallel geometry lowers the decay exponent $\mu$, which gradually becomes smaller than unity (see Fig~9 in Ref.~\cite{Dwivedi}) and we claimed that this effect is associated with the fact that under rotation thermal counterflow acquires two-dimensional features.

The effect of horizontal rotation, $\mathbf{\Omega} \perp \mathbf{v}_{\rm{ns}}$, on the temporal decay $L(t, \Omega)$ seems  more complex. Here the slow rotation slightly lowers the decay exponent $\mu$, too, but for faster rotation rates the decay gradually ceases to display clear power law. Instead, the faster decay occurs, displays rather rounded decay curves which become gradually shifted toward shorter and shorter times. 

\section{Discussion}

Direct comparison of experimental results obtained in $\mathbf{\Omega} \perp \mathbf{v}_{\rm{ns}}$ and $\mathbf{\Omega} \parallel \mathbf{v}_{\rm{ns}}$ geometries of rotating thermal counterflow in He~II shows that there are no appreciable effects in their steady states, in that the directly measured vortex line density agrees within experimental uncertainty with each other, with possible delay in the onset of turbulence in the perpendicular case. It is hardly surprising, as the turbulence-generated vorticity vastly overwhelms the rotation-induced vortex line density. On the other hand, the transient behavior in $\mathbf{\Omega} \perp \mathbf{v}_{\rm{ns}}$ and $\mathbf{\Omega} \parallel \mathbf{v}_{\rm{ns}}$ geometries differ significantly. 

To this end, we find it useful to discuss the complementary decays of classical grid-generated turbulence. Classical 3D turbulence under~rotation acquires two-dimensional (2D) features.  A prime example is the temporal decay of energy per unit mass of the grid generated nearly homogeneous and isotropic turbulence which, without~rotation (past initial transition stage), is at late times well described by a power law $E(t)=u^2(t) \propto (t-t_{\rm{vo}})^{-n}$, where $n$ is the decay exponent and $t_{\rm{vo}}$ denotes a virtual origin, i.e., the hypothetical time instant when the turbulent energy decaying according this power law would have been infinite. The~value of the exponent $n$ depends on whether the size of the energy containing
eddies is free to grow ($n \approx 6/5$) or is bounded by the domain size ($n \approx 2$). For~decaying vorticity $\omega$ this transfers, using the relationship $\varepsilon=-dE/dt=\nu \omega ^2$ which follows from the Navier-Stokes equation, to power law decay with exponents $\mu \approx 11/10$ and $\mu \approx 3/2$. We note in passing that these decay exponents are, via relationship $\omega=\kappa L$, applicable to decaying vortex line density in quasi-classical quantum turbulence of the Kolmogorov form \cite{Stalp1999,Skrbek2000}. 

Morize and Moisy~\cite{Morize2006} studied the energy decay of grid-generated turbulence in a rotating tank by means of particle image velocimetry. They used a water filled glass tank of square section, 35 cm on the side, mounted on a rotating turntable, whose angular velocity $\Omega$ was varied between 0.13 and 4.34 rad/s.  After~the fluid was set in solid body rotation, turbulence was generated by rapidly towing a co-rotating square grid of square bars of 1~cm with a mesh size of $M =39$ mm. The~authors discussed three characteristic times: in addition to the instantaneous turnover time $M/u$, where $M \approx \ell_{\rm{ece}}$ is the characteristic size of the energy-containing eddies, two other time scales are present in the problem, which have opposite effects on the turbulence decay. One is the rotation time scale $T_{\rm{rot}}=\Omega^{-1}$ and, for~bounded systems, there is also the Ekman time scale, $T_{\rm{Ek}}=H (\nu \Omega)^{-1/2}$, where $H$ is the characteristic size of the turbulent system along the rotation axis. The~rotation time scale is associated with the propagation of inertial waves, which modify the nonlinear energy transfer and reduce the energy dissipation rate. This results in a lower value of the decay exponent $n$. Contrary to that, the~Ekman time scale governs the dissipation of those inertial waves from multiple reflections in the Ekman layers, thus enhancing the energy decay at large times, and~shortening the range for a possible self-similar decay even at large Reynolds numbers. Are these considerations applicable for decaying rotating counterflow turbulence?

We claim that our complementary experiments performed in $\mathbf{\Omega} \perp \mathbf{v}_{\rm{ns}}$ and $\mathbf{\Omega}  \parallel \mathbf{v}_{\rm{ns}}$ geometries of rotating thermal counterflow confirm both effects. Let us define the effective Ekman time as $T_{\rm{Ek}}^{\rm{eff}}=H (\nu_{{\rm{eff}}} \Omega)^{-1/2}$, where $\nu_{\rm{eff}}$ is the effective kinematic viscosity of turbulent He~II. This weakly temperature dependent quantity has been introduced, measured and discussed in a number of experiments over almost entire temperature range from $T_\lambda$ down to zero temperature limit  \cite{QTbook,ChagovetsGorSkr,Zmeev2015,Gao2018} and is estimated to reach $0.1 \kappa \lesssim \nu_{\rm{eff}} \lesssim \kappa \approx 10^{-7}$m$^2$s$^{-1}$.
The effective Ekman time $T_{\rm{Ek}}^{\rm{eff}}$, proportional to the size of the system along the rotation axis, is due to the geometry of the counterflow channel for the $\mathbf{\Omega} \perp \mathbf{v}_{\rm{ns}}$ geometry about 12 times shorter than for the $\mathbf{\Omega} \parallel \mathbf{v}_{\rm{ns}}$ geometry. Here, even for the highest $\Omega \approx \pi/3$ explored \cite{Dwivedi}, the effective Ekman time exceeds about a minute, so over more than a decade in time one experimentally observes  a well-pronounced power law decays of the form  $L(t) \propto (t-t_{\rm{vo}})^{-\mu}$. The corresponding decay exponents $\mu$ are shown in Fig.~9 of Ref.~\cite{Dwivedi}. As shown in Fig.~\ref{fig:decayCompar}, in the $\mathbf{\Omega} \perp \mathbf{v}_{\rm{ns}}$ case the effective Ekman time reaches, in dependence of the rotation rate, seconds and affects the temporal decay $L(t,\Omega)$ appreciably. This prevents reliable extracting of the decay exponents, except perhaps for rough estimate of $1.1 \lesssim \mu \lesssim 1.2$ for low rotation rate of 60 deg/s, which qualitatively confirms the lowering tendency of $\mu$ with increasing rotation rate.

As recently suggested in Ref.~\cite{SkrbekEntropy}, it will be interesting to investigate the decay of rotating grid-generated quantum turbulence in He~II, for which the analogy with the decay of bounded classical grid turbulence \cite{Morize2006,Bewley2007} is more straightforward. 

\section{Conclusions}

 By using the second sound attenuation technique we have studied horizontally ($\mathbf{\Omega} \perp \mathbf{v}_{\rm{ns}}$) and axially ($\mathbf{\Omega} \parallel \mathbf{v}_{\rm{ns}}$) rotating thermal counterflow of superfluid $^4$He (He~II) generated thermally in a square channel. In both geometries, the array of rectilinear quantized vortices created in the channel by rotation strongly affects the development \cite{Dwivedi}, steady state and temporal decay of quantum turbulence, which under rotation acquires two-dimensional features. While there are no appreciable effects in their steady states, the transient behavior in $\mathbf{\Omega} \perp \mathbf{v}_{\rm{ns}}$ and $\mathbf{\Omega} \parallel \mathbf{v}_{\rm{ns}}$ geometries differ significantly, being affected by three characteristic time scales, namely by instantaneous turnover time of the largest eddy, the rotation time scale $T_{\rm{rot}}=\Omega^{-1}$ and the effective Ekman time scale $T_{\rm{Ek}}^{\rm{eff}}=H (\nu_{{\rm{eff}}} \Omega)^{-1/2}$, proportional to the size of the system along the rotation axis. In particular, two effects are observed, acting against each other and affecting the late temporal decay of vortex line density. The first is gradual decrease of the decay exponent, clearly observed in the $\mathbf{\Omega} \parallel \mathbf{v}_{\rm{ns}}$ geometry \cite{Dwivedi}. It exists in the $\mathbf{\Omega} \perp \mathbf{v}_{\rm{ns}}$ geometry, too, but it is screened here by the influence of the effective Ekman layer built in the turbulent superfluid He~II, significantly shortening the time range for a possible self-similar decay of vortex line density.

 We believe that our findings will stimulate further studies of quantum flows of He~II in order to improve our understanding of the underlying physics of superfluid hydrodynamics.

\section*{Acknowledgements}
 Fruitful discussions with K.R. Sreenivasan are appreciated. This research was supported by Czech Science Foundation (GACR) under \#25-16588S.

\bibliography{rotating-counterflow}

\end{document}